# Raman LIDARs for the atmospheric calibration along the line-of-sight of CTA

**Otger Ballester,**[a] **Oscar Blanch,**[a] **Joan Boix,**[a] **Johan Bregeon,**[b] **Patrick Brun,**[b] **Şıdıka Merve Çolak,**[a] **Michele Doro,**[c] **Vania Da Deppo,**[d] **Lluís Font,**[e] **Omar Gabella,**[b] **Rafael García,**[a] **Markus Gaug,**[e] **Camilla Maggio,**[e] **Manel Martínez,**[a] **Òscar Martínez,**[a] **Pere Munar Adrover,**[e] **Raul Ramos,**[e] **Stephane Rivoire,**[b] **Samo Stanič,**[f] **David Villar,**[e] **George Vasileiadis**[*,b] **Longlong Wang,**[f] **Marko Zavrtanik,**[f,g]

[a] *Institut de Fisica d'Altes Energies (IFAE), The Barcelona Institute of Science and Technology, Campus UAB, 08193 Bellaterra (Barcelona), Spain*

[b] *Laboratoire Univers et Particules de Montpellier, Université de Montpellier, CNRS/IN2P3, France*

[c] *INFN Sezione di Padova and Università degli Studi di Padova, Via Marzolo 8, 35131 Padova, Italy*

[d] *CNR-Institute for Photonics and Nanotechnologies UOS Padova LUXOR, Via Trasea 7, 35131 Padova, Italy*

[e] *Unitat de Física de les Radiacions, Departament de Física, and CERES-IEEC, Universitat Autònoma de Barcelona, E-08193 Bellaterra, Spain*

[f] *Center for Astrophysics and Cosmology, University of Nova Gorica, Vipavska 11c, 5270 Ajdovščina, Slovenia*

[g] *Jožef Stefan Institute, Jamova 39, 1000 Ljubljana, Slovenia*

The Cherenkov Telescope Array (CTA) is the next generation ground-based observatory for gamma-ray astronomy at very-high energies. Employing more than 100 Imaging Atmospheric Cherenkov Telescopes in the northern and southern hemispheres, it was designed to reach unprecedented sensitivity and energy resolution. Understanding and correcting for systematic biases on the absolute energy scale and instrument response functions will be a crucial issue for the performance of CTA. The LUPM group and the Spanish/Italian/Slovenian collaboration are currently building two Raman LIDAR prototypes for the online atmospheric calibration along the line-of-sight of the CTA. Requirements for such a solution include the ability to characterize aerosol extinction at two wavelengths to distances of 30 km with an accuracy better than 5%, within time scales of about a minute, steering capabilities and close interaction with the CTA array control and data acquisition system as well as other auxiliary instruments. Our Raman LIDARs have design features that make them different from those used in atmospheric science and are characterized by large collecting mirrors ($\sim$2.5 m$^2$), liquid light-guides that collect the light at the focal plane and transport it to the readout system, reduced acquisition time and highly precise Raman spectrometers. The Raman LIDARs will participate in a cross-calibration and characterization campaign of the atmosphere at the CTA North site at La Palma, together with other site characterization instruments. After a one-year test period there, an in-depth evaluation of the solutions adopted by the two projects will lead to a final Raman LIDAR design proposal for both CTA sites.

*36th International Cosmic Ray Conference -ICRC2019-*
*July 24th - August 1st, 2019*
*Madison, WI, U.S.A.*

---

*Speaker.

## 1. Introduction

The CTA Atmospheric Calibration Strategy [1] relies on the characterization of the atmospheric aerosol profile along the line-of-sight of the observing CTA array with the help of a powerful Raman LIDAR, and the follow-up characterization of the observed field-of-view with the help of a wide-field stellar photometer [2]. Both instruments are necessary to obtain a complete, accurate, timely and wavelenth-resolved aerosol extinction model for the Cherenkov light of the gamma-ray air showers observed by CTA's Imaging Atmospheric Cherenkov Telescopes (IACTs) [3].

## 2. Main requirements for the CTA Raman LIDAR

Given the typical altitudes of Extended Air Showers, from which CTA will collect Cherenkov light, the aerosol transmission profile along the line-of-sight of CTA must be monitored up to altitudes of 15 km a.s.l. This leads to a required LIDAR range of 30 km at least, once the full zenith angle range of the CTA Observatory is taken into account. The typical dimensions of the observed air showers, ranging over several kilometers suggest that the retrieved transmission profile should have a range resolution of at least or better than $\sim$300 m. Moreover, since the analyzed Cherenkov light shows photon energies within a range of about a factor of two, the atmosphere shall be characterized with at least two wavelengths representative for that Cherenkov spectrum. Finally, the Cherenkov light transmission probability due to the presence of aerosols must then be measured with an absolute accuracy of 0.03 at least, to meet the strict CTA requirements on the accuracy with which the Cherenkov photon density gets measured on ground.

Early studies [4] have already shown that the aerosol transmission profiling is best determined with powerful Raman LIDARs. Such LIDARs should be equipped with near-range optics, in order to determine the full ground layer transmission reliably and such meet the aerosol transmission accuracy requirement. Moreover, stratospheric aerosol extinction should be accessible to the LIDAR, at least when pointing towards the zenith.

LIDARs can operate with frequency doubled and tripled Nd:YAG laser light, hence characterizing the atmosphere at 355 nm and 532 nm, about the right wavelength choice to cover the measured Cherenkov light spectrum. Because the CTA telescopes will be blinded by a LIDAR shooting into their field-of-view with these wavelengths at high intensity, the LIDAR must permit to fully characterize one profile a few minutes before and after a CTA observation period, and during the change of Wobble position, in the latter case *during a time interval of about a minute*.

## 3. Common parts and concepts

To achieve the required long ranges, an unusually large mirror of 1.8 m diameter is used, and the system equipped with a powerful laser. The designs shown in this proceeding have refurbished each a telescope of the discontinued CLUE experiment [8].

The outgoing laser beam is directed towards the telescope pointing axis in a co-axial configuration, ensuring full overlap at little more than hundred meters.

The 1.8 m mirror is parabolic, with an $f$-number of 1, the whole located inside a standard shipping container whose two side had been modified to become openable doors.

For this design, a so-called *link-budget* calculation [6] had been performed showing that the dimmest Raman line will be detected from a distance of 13 km (i.e. altitude about 15 km a.s.l.) with a signal-to-noise ratio of 10 after only one minute. Most of the numbers derived in [6] have been recently validated with first-light data of the Barcelona Raman LIDAR [7].

## 4. The Montpellier Design

The LUPM Raman version uses a Quantel CFR 400 Nd:YAG laser as a UV source generator. It is firmly mounted on the top of the telescope's dish structure. It is a military grade 10 ns pulsed laser, 4.2 W total power, with a repetition rate of 20 Hz, providing three different wavelengths at 1064 nm, 532 nm and 355 nm. The laser beam exits the telescope dish using a double UV coated mirror guiding system. To achieve optimal performance, a precise alignment system is needed to assure that at any moment the optical axis of the telescope and the laser beam are co-linear. We have opted for a computerized industrial system by ThorLabs, with a step size precision of less than 1 mm. The light output from the focal point of the telescope is fed to the entrance of the Raman polychromator via a Lumatec 300 series liquid fiber of 8 mm diameter with a numerical aperture of 0.59. A support for the light guide has also been installed in the structure of the telescope. All parts are fixed into this structure in such a way that movements are restricted and a minimum bending radius for the liquid fiber light guide is assured.

Figure 1 presents the polychromator built by the Raymetrics company. It is is designed to be lightweight, mechanically modular, and optically efficient. Dichroic mirrors are used to separate the different Raman and elastic lines, while a dedicated two-lens eye piece is used in front of every PMT to focus uniformly the incident light. Custom 2 inch optics are used throughout the polychromator to match efficiently the PMT entrance window size.

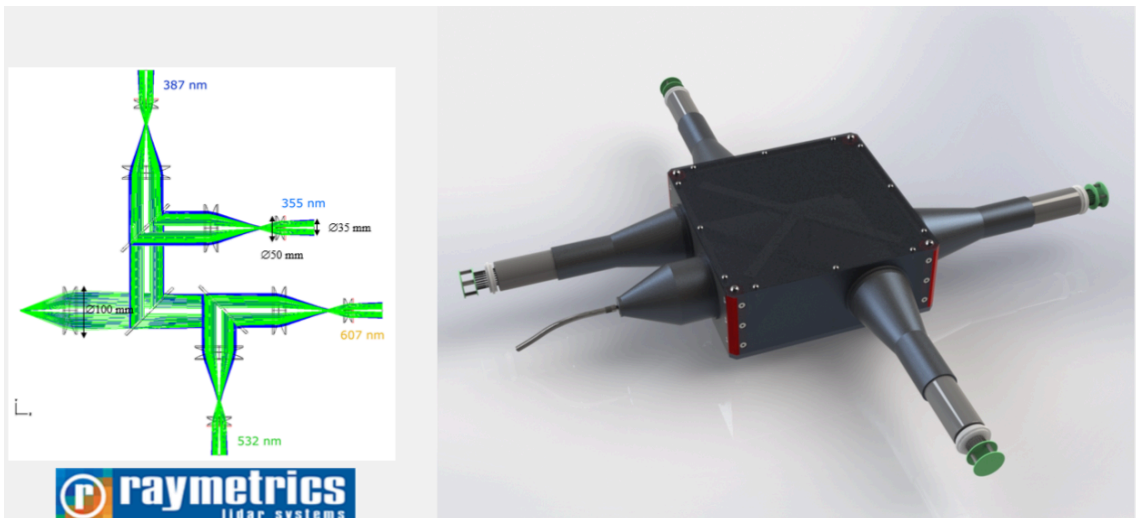

**Figure 1:** On the left side the *Zemax* design of the polychromator is presented. At the right side the final mechanical CAD design under production. All images provided by Raymetrics.

The polychromator is now finalized, and its integration will happen soon in Montpellier.

## 5. The Barcelona Design

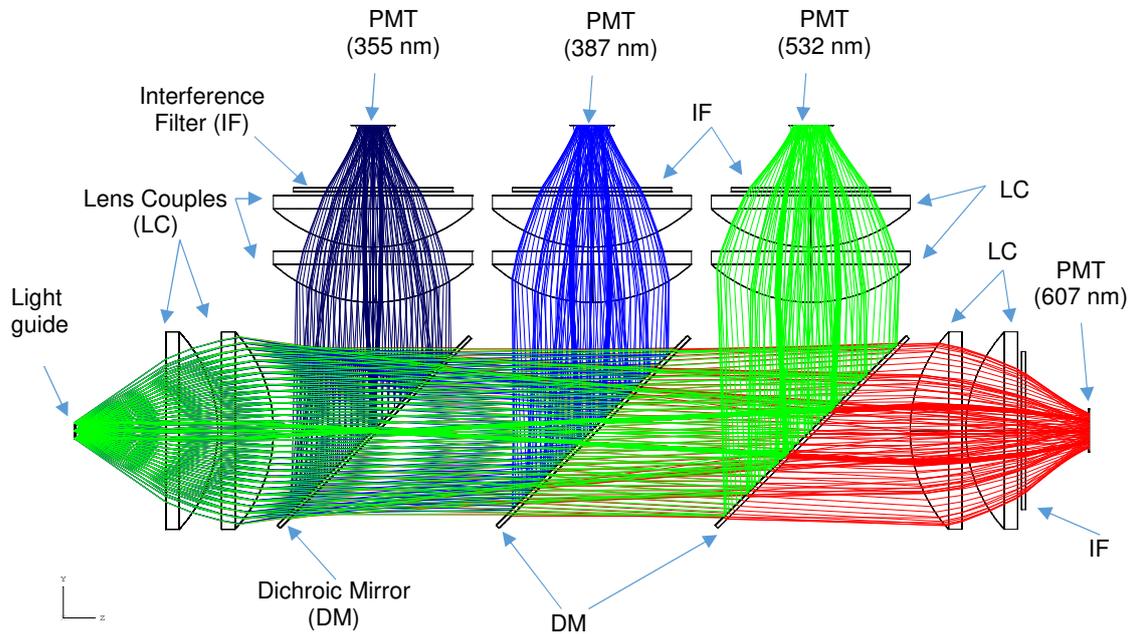

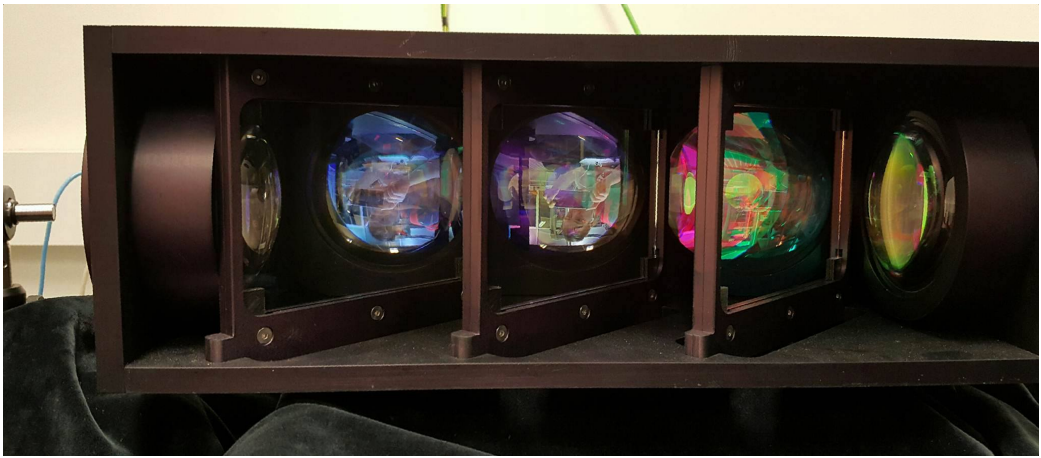

**Figure 2:** Top: the *Zemax* design of the Barcelona LIDAR polychromator. Bottom: A picture of the polychromator taken from the bottom side.

The Barcelona design uses a Nd:YAG laser from *Quantel Brilliant* with a frequency doubling and tripling head, a pulse of 5 ns duration and energy of ∼250 mJ and a repetition rate of 10 Hz. The laser operates at the wavelengths 355 nm, 532 nm and 1064 nm and is mounted on a millimetric *xy*-adjustable arm. As shown in Fig.x [Fig.], the laser light is reflected by two dichroic steering mirrors, used to achieve co-axial configuration and to absorb the 1064 nm wavelength (which might damage the later introduced liquid light guide). The back-scattered light is collected by the main mirror and focused on an 8 mm diameter liquid light guide (LLG) of type *Lumatec Series 300*, which has been characterized thoroughly for its transmission properties under different input and output angles and during vibration. The collected light optimally transferred to an in-house built polychromator box, mounted on the back side structure of the main mirror. That 760×550×170 mm sized, 30 kg heavy box of anodized aluminum separates the incoming light successfully into two elastic (355 nm, 532 nm) and the two Raman (387 nm, 607 nm) wavelengths.

The optics of the polychromator (see Fig. 2) has been designed in collaboration with the CNR Institute for Photonics and Nanotechnologies in Padova, Italy [9]. The mechanical design and tests were carried out at IFAE. After collimation of by 100 mm Lens Couples (LC), the incoming light is directed towards its respective detector with the help of three Dichroic Mirrors (DM). In each channel, the light is again focused by LCs and noise-reduced by 10 nm wide Interference Filters (IFs). Finally, each of the four wavelengths is collected by a 1.5 inch, high quantum efficiency PMT of type Hamamatsu R11920, the same as those used for the Large-Sized-Telescope camera [11].

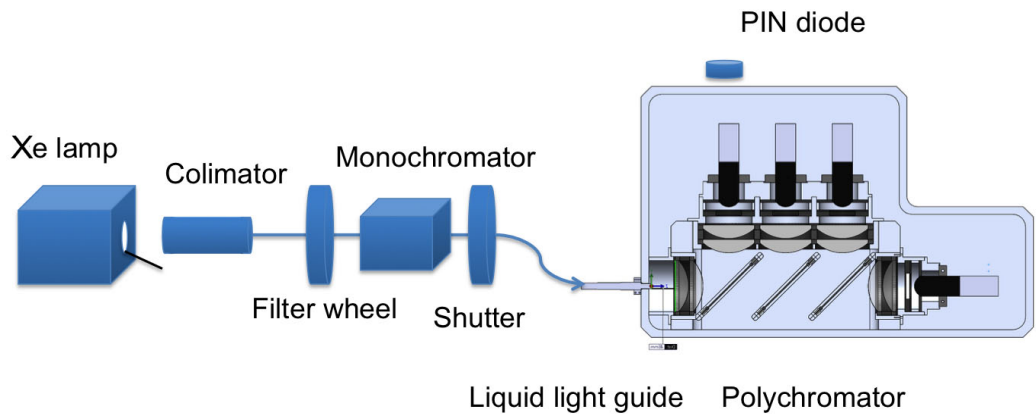

**Figure 3:** A sketch of the measurement setup to characterize the polychromator.

After all the individual characterization tests of every pieces, light leakage tests were carried out with the setup shown in Fig. 3. Light produced by a stabilized Xe-lamp passes through a collimator and a set of filters. Wavelengths selected by a monochromator are transmitted to the polychromator through an automatic shutter and the LLG. Ten thousand of measurements were taken for each wavelength in open and closed mode in order to subtract the background and to achieve statistically significant results. Light leakage from outside the polychromator box and between the channels were tested. Leakage from the elastic into the Raman channel could be excluded to be larger than $2 \times 10^{-7}$.

## 5.1 First Results

The Barcelona LIDAR has seen its first light in July 2018 [7] and has undergone several minor optimisations and modifications since then. Back-scattering signals in the two elastic channels (at 355 nm and 532 nm) and one Raman channel (387 nm) were detected by the corresponding photomultipliers (PMTs) with an applied voltage of 1500 V, using the fully-powered laser as the transmitter (see Figure 4). Signals were amplified and digitized using both analog and photon counting (PC) detection chains implemented in the Licel transient recorders [12]. For the PC chain, the raw signal was used after a dead-time correction (6.2 ns) of the PMTs. After range correction and noise subtraction, signals from both detection chains were glued by adjusting the analog trace following [13] in order to improve the detection range and the signal to noise ratio. More elaborate data gluing methods, e.g. from [14] are currently being implemented, however the shown results already show satisfactory overlap of the both readout modes across more than 5 km range for the elastic channels and 2 km range for the Raman channel. The collected data are based on an average of 500 shots within 50 s. The detectable range (signal to noise ratio larger than **1**) in the elastic channels was 25 km, while the detectable range for the Raman signal at 387 nm reached only about 10 km (Figure 4). In the presented example, the signals from elastic channels (355 nm and 532 nm) were saturated in the near range (below 1.5 km and 2 km respectively) in the amplification and digitalisation stage due to the high aerosol loading and large PMT signals. In order to reduce the saturation in the near range, PMT gains of the elastic channel were reduced (the voltage was decreased to ∼800 V), however, this strongly affected far range detection and thus decreased the detectable range (see Figure 5). Due to the weak signal in the far range, filter methods [15] together with the Raman inversion [16] were tested only in the range of good signal to noise ratio of the analog readout chain.

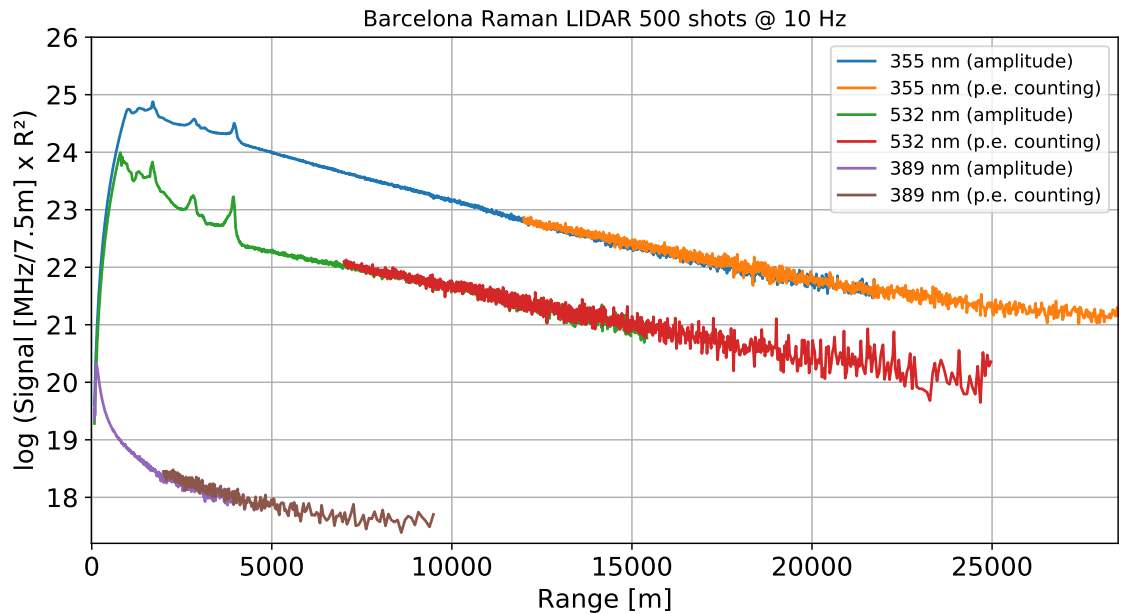

**Figure 4:** First light range-corrected signals from the three colour lines of the Barcelona Raman LIDAR (two elastic and one Raman channel), based on 50 s (500 laser shots) of oversampling. The analog (amplitude) and photo-electron (p.e.) counting parts are shown where applicable. The features visible in the elastic lines below 4 km correspond to aerosol layers and clouds.

The detectable range of this system was found to be seriously affected by the reflectance of the primary mirror, which in the test case caused about 50% loss of emitted laser. In future operation, this problem will be resolved by cleaning the mirror.

## 6. Conclusions

The two Raman LIDAR prototypes for CTA are in good shape and begin to demonstrate their expected performance. In Barcelona, further easy-to-achieve improvements concern a gen-eral cleaning of the mirror, expected to significantly increase its reflectance, and to increase the laser rate from currently 10 Hz to 20 Hz or even 50 Hz. Moreover, we concentrate on further robotisation of the system and to achieve compliance with all safety and engineering requirements as well as the communication protocols used by CTA. Finally, a dedicated near-range readout for one elastic channel is currently being commissioned and expected to improve the accessible range of the Raman LIDAR down to almost 30 m distance. Once all these steps are completed, the Barcelona LIDAR will undergo a one-year test at the Observatorio del Roque de los Muchachos, participating in a cross-calibration campaign with an EARLINET[1]-calibrated LIDAR for site characterization [2]. Experience has shown that Raman LIDARs and their data analysis methods should be validated through recognized LIDAR networks, such as EARLINET in Europe.

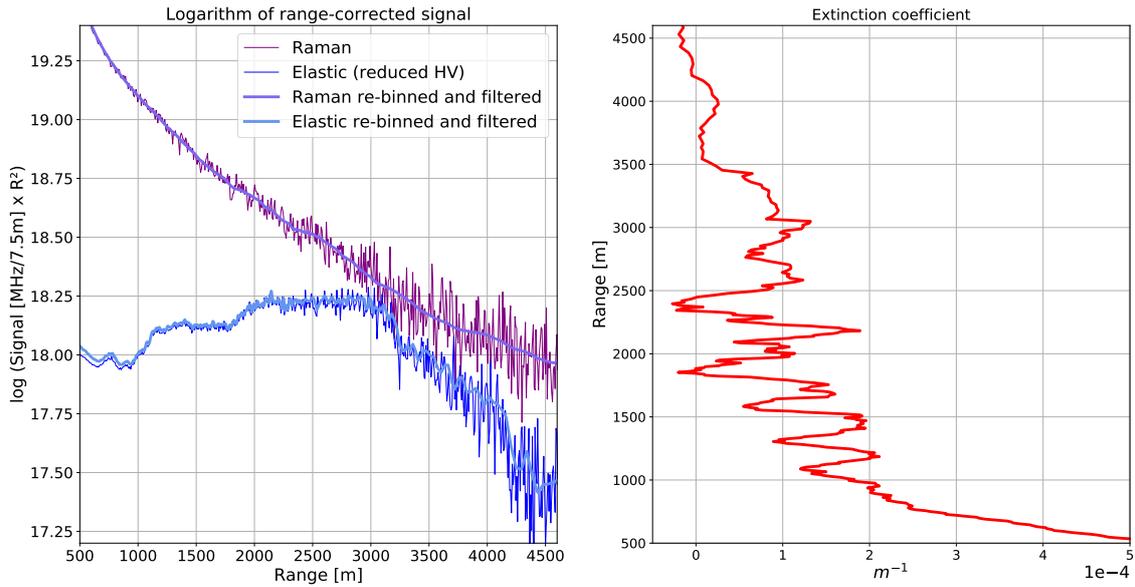

**Figure 5:** Left: Range-corrected signal from a test run with one Raman channel and one elastic channel using PMTs operating at a reduced voltage of 800 V. Right: Preliminary analysis of the inverted Raman signal yielding atmospheric extinction profile above UAB.

Both Montpellier and Barcelona LIDAR designs will be thoroughly compared and finally harmonized by standardizing the respective better performing components, before they can be delivered as in-kind-contributions to CTA.

---

[1]https://www.earlinet.org/